\documentclass[superscriptaddress,nofootinbib,aps,prd,floatfix]{revtex4}
\pdfoutput=1
\usepackage{amsmath,multirow,graphicx,tabularx,epsfig}
\usepackage{color}
\usepackage{hyperref}
 
\newcommand\one{\leavevmode\hbox{\small1\normalsize\kern-.33em1}}

\newcommand{\qqqquad}{\qquad \qquad \qquad}

\newcommand{\gev}{{\ensuremath\rm GeV}}

\newcommand{\ifb}{{\ensuremath\rm fb^{-1}}}

\def\slashchar#1{\setbox0=\hbox{$#1$}           
   \dimen0=\wd0                                 
   \setbox1=\hbox{/} \dimen1=\wd1               
   \ifdim\dimen0>\dimen1                        
      \rlap{\hbox to \dimen0{\hfil/\hfil}}      
      #1                                        
   \else                                        
      \rlap{\hbox to \dimen1{\hfil$#1$\hfil}}   
      /                                         
   \fi}

\def\eg{{\sl e.g.} \,}
\def\ie{{\sl i.e.} \,}

\begin{document}

\title{MadMax, or Where Boosted Significances Come From}

\author{Tilman Plehn}
\affiliation{Institut f\"ur Theoretische Physik,
             Universit\"at Heidelberg, Germany}
\author{Peter Schichtel}
\affiliation{Institut f\"ur Theoretische Physik,
             Universit\"at Heidelberg, Germany}
\author{Daniel Wiegand}
\affiliation{Institut f\"ur Theoretische Physik,
             Universit\"at Heidelberg, Germany}
\affiliation{PITT-PACC, Department of Physics and Astronomy, University of Pittsburgh, US}

\begin{abstract}
In an era of increasingly advanced experimental analysis techniques it
is crucial to understand which phase space regions contribute a signal
extraction from backgrounds. Based on the Neyman-Pearson lemma we
compute the maximum significance for a signal extraction as an
integral over phase space regions. We then study to what degree
boosted Higgs strategies benefit $ZH$ and $t\bar{t}H$ searches and
which transverse momenta of the Higgs are most promising. We find that
Higgs and top taggers are the appropriate tools, but would profit from
a targeted optimization towards smaller transverse momenta. MadMax is
available as an add-on to Madgraph5.\end{abstract}
 
\maketitle
 
\bigskip
\bigskip
\bigskip

\tableofcontents

\newpage

\section{Maximum significance}
\label{sec:intro}

The recent Higgs discovery has shown that modern analysis techniques
have become standard in high energy physics. Such techniques go beyond
simple event counting in phase space regions which have been
identified as signal--rich ahead of time.  Multi-variate strategies
with ten or more kinematic observables seem to make it impossible for
the experimental collaborations to provide sufficient information on
the behavior of each individual observable and their correlation in
the signal and background phase spaces. This means that our research
field would benefit from a compact tool to study the leading effects
in the computation of quoted significances from multi--variate
analyses~\cite{orig}.

More specifically, when analyses like Higgs coupling measurements
start to be limited by theory uncertainties we need to clearly
identify the phase space regions which carry the analysis
result~\cite{sfitter}. Moreover, recent progress in boosted Higgs and
top studies~\cite{bdrs,tth,buckets} has shown that identifying the
appropriate phase space patterns can trigger the development of
entirely new, specialized analysis objects like fat
jets~\cite{top_tagger} or deconstructed parton
showers~\cite{deconstruction}. Again, this points to the need of a
fast Monte--Carlo tool which can reliably identify those phase space
regimes which are critical to separating a given signal from
backgrounds~\cite{orig}.\medskip

In \textsc{MadMax} we will not attempt include all detector effects,
because matrix element techniques including appropriate transfer
functions are hugely computer intensive. For the same reason, we will
limit ourselves to the parton level, assuming that the key phase space
patterns of signal and background processes are defined by the hard
processes. This allows us to use the \textsc{Madgraph}~\cite{madgraph}
framework for most of our event generation. On the other hand, these
two approximations should have a clearly defined mathematical effect
on the bottom line, in our case stating that our result gives an {\sl
  upper limit} on the significance any full analysis can reach.

The setup of such a tool has been developed for the example process of
Higgs production in weak boson fusion with a subsequent decay to muons
many years ago~\cite{orig}. It allows us to compute the maximal
significance with which we can separate a signal--plus--background
hypothesis from the background only based on Monte--Carlo event
generation. It has the key feature that it allows for cuts on the
contributing phase space and computes this maximum significance as a
strictly increasing function when we add more phase space regions. We
will rely on this feature to answer the key physics question of this
paper: {\sl how much boost should we target in boosted Higgs and top
  searches?}\medskip

Mathematically, our computation is inspired by the Neyman--Pearson
lemma, stating that the likelihood ratio is the most powerful variable
to distinguish between a background hypothesis and the
signal--plus--background hypothesis~\cite{proof}. This is formally
defined as the minimum probability for a false negative outcome given
a fixed probability for the false positive signal outcome. If we
assume that the signal--plus--background hypothesis is true this
implies the lowest probability of mistaking the signal for a
background fluctuation.

In experiment we have to measure any multi-dimensional probability
density function. In our approach we use the parton--level transition
amplitude for signal and background processes to compute the
probability density over the full phase space at a given order in
perturbation theory~\cite{early_mem}. We use a similar notation to the
so-called matrix element method~\cite{mem}, but emphasize that within
\textsc{Madgraph} the experimental matrix element approach is already
supported by \textsc{MadWeight}~\cite{madweight}.\medskip

For now limiting ourselves to irreducible backgrounds, \ie signal
and background processes with identical degrees of freedom in the
final state, we can simultaneously probe the signal and background
phase space using a vector of random numbers $r$, with or without
acceptance cuts,
\begin{equation}
\sigma_\text{tot} 
= \int_\text{cuts} dr \; M(r) \; d\sigma(r) \; .
\label{eq:int_first}
\end{equation}
The phase space boundaries are included in the integral, and the
differential cross section $d\sigma(r)$ includes all phase space
factors and the Jacobian for transforming the integration to the
random number basis. The integration over the parton distribution
momentum fractions $x_{1,2}$ is included in the phase space integral.
A measurement function $M$ can parameterize additional cuts or detector
efficiencies. Because $r$ is a basis vector, cuts on observable
quantities consistently remove these phase space regions from all
processes. All potentially available information is included in the
array of event weights $M(r) \, d\sigma(r)$.\medskip

A cut analysis defines a signal--rich region and then counts events in
that region.  The variable that discriminates between signal and
background is the number of events ($s,b$) in this region. For
counting analyses the likelihood of observing $n$ events assuming the
background-only hypothesis is given by the Poisson distribution
$\text{Pois}(n|b)=e^{-b} \, b^n/n!$. We can generalize this number
counting by introducing a discriminating observables vector $x$. We
assume that the background--only hypothesis $H_b$ is described by the
normalized distribution $f_b(x)$, while the signal--plus--background
hypothesis $H_{s+b}$, assuming no interference, is described by
$f_{s+b}(x) =[sf_s(x) + bf_b(x)]/(s+b)$. Following the Neyman--Pearson
lemma, the most powerful test statistic is the likelihood ratio. The
total likelihood for the full vector $x=\{x_j\}$ can be factorized
into the Poisson likelihood to observe $n$ events, and the product of
the individual event's likelihood $f(x_j)$,
\begin{alignat}{5}
q(x) &= \log \; \frac{L(x|H_s)}{L(x|H_b)}
      = \log \; \frac{\text{Pois}(n|s+b)  \; \prod_{j=1}^n f_{s+b}(x_j)} 
             {\text{Pois}(n|b)    \; \prod_{j=1}^n f_b(x_j)} \notag \\
     &= \log \left[ e^{-s} \; \left( \frac{s+b}{b} \right)^n \;
                    \frac{\prod_{j=1}^n f_{s+b}(x_j)}
                         {\prod_{j=1}^n f_b(x_j)}
            \right] \notag \\
     &= {-s}  +  \sum_{j=1}^{n} \log \left( 1+\frac{s f_s(x_j)}
                                                       {b f_b(x_j)}
                                                 \right) \; .
\label{eq:llr}
\end{alignat}
The key step of our description is to generalize the observables
vector $x$ to all individual phase space points $r$. Following
Eq.\eqref{eq:int_first} they are probed by the Monte Carlo generation,
which means that we can compute the normalized probability
distributions $f(x)$ from the parton--level matrix elements and
construct a log--likelihood ratio map of all final state
configurations using the normalized probability distributions
$d\sigma(r)/\sigma_\text{tot}$ for the signal and background,
\begin{equation}
q(r) = -\sigma_{\text{tot},s} \; \mathcal{L} \; + \,
             \log \left( 1 + \frac{d\sigma_s(r)}
                                 {d\sigma_b(r)}  
                 \right) \; .
\label{eq:likelimap}
\end{equation}
${\cal L}$ denotes the integrated luminosity. To construct the
single--event probability distribution $\rho_{1,b}(q)$ we combine the
background event weight with the log--likelihood ratio map $q(r)$,
\begin{equation}
\rho_{1,b}(q_0) = \int dr \; \frac{d\sigma_b(r)}{\sigma_{\text{tot},b}}
                         \; \delta \left( q(r) - q_0 \right) \; .
\label{eq:rho_1}
\end{equation} 
For multiple events, the distribution of the log--likelihood ratio
$\rho_{n,b}$ can be computed by repeated convolutions of the single
event distribution. This convolution we can evaluate using a Fourier
transform~\cite{clfft}.  The expected log--likelihood ratio
distribution for a background including Poisson fluctuations in the
number of events $n$ is
\begin{equation}
\rho_b(q) = \sum_n \text{Pois}(n|b) \times \rho_{n,b}(q) \; . 
\label{eq:rho_b}
\end{equation} 
To compute it from the single--event likelihood $\rho_{1,b}(q)$
we first Fourier transform all functions $\rho(q)$ into complex--valued
functions of the Fourier--transformed likelihood ratio
$\overline{\rho_{1,b}}(\overline{q})$. The convolution in $q$ space
becomes a multiplication in Fourier space, namely
$\overline{\rho_{n,b}} = (\overline{\rho_{1,b}})^n$.  The sum over $n$
in Eq.\eqref{eq:rho_b} has the closed form $\overline{\rho_b} = \exp[b
  \; (\overline{\rho_{1,b}} - 1)]$.  For the signal--plus--background
hypothesis we expect $s$ events from the $\rho_{1,s}$ distribution and
$b$ events from the $\rho_{1,b}$ distribution. Similar to the above
formula we have $\overline{\rho_{s+b}} = \exp[ b
  (\overline{\rho_{1,b}} - 1) + s (\overline{\rho_{1,s}} - 1)]$. A
transformation back into $q$ space gives us log--likelihood ratio
distributions $\rho_b(q)$ and
$\rho_{s+b}(q)$~\cite{lepstats}. Finally, given a value $q$ we can
calculate the background--only confidence level
\begin{equation}
\text{CL}_b(q) =\int_{q}^\infty dq' \; \rho_b(q') \; .
\label{eq:clb}
\end{equation}
To estimate the discovery potential of a future experiment we assume
the signal--plus--background hypothesis to be true and compute $\text{
  CL}_b$ for the median of the signal--plus--background distribution
$q^*_{s+b}$.  This expected background confidence level can be
converted into an equivalent number of $Z$ Gaussian standard
deviations by implicitly solving $\text{CL}_b(q^*_{s+b}) = \left(
1-\text{erf}(Z/\sqrt{2}) \right)/2$.\medskip

In general it is clear how to include detector effects in our
simulation. However, to determine the maximal significance in a strict
sense we should not include detector effects, because they always
decrease the significance. In our case lepton and jet directions are
usually well measured. The jet energy scale can be an issue for the
detailed analysis, but we do not expect it to have a great effect on
our results, either. Combinatorics will eventually be an issue, but
again it will not be critical for the analyses we present in this
paper. In contrast, the experimental resolution of $m_{bb}$ is nowhere
close to the physical Higgs width in the Standard Model. We therefore
introduce a Gaussian smearing for this one observable. The convolution
of the physical Higgs width with this Gaussian we can safely
approximate as the Gaussian detector resolution
alone~\cite{orig}.\medskip

We will discuss two specific analyses in this paper, boosted Higgs
searches in $ZH$ production~\cite{bdrs} and in $t\bar{t}H$
production~\cite{tth}. They are crucial for a model--independent
determination of the heavy quark Yukawa couplings at the LHC, \ie for
the measurement of the most sensitive probes for new physics in the
Higgs sector~\cite{2hdm,lecture}.  In both cases we limit ourselves to
the irreducible backgrounds, \ie the processes where the $H \to
b\bar{b}$ decay is replaced by $Z \to b\bar{b}$ and QCD $b \bar{b}$
continuum production. The additional final state particles we assume
to be fully reconstructed. For the case of the $Z$ boson discussed in
Section~\ref{sec:zh} this is clearly realistic, as long as we rely on
leptonic $Z$ decays. Since our analysis focuses on the kinematics on
the $b\bar{b}$ system our findings can be generalized to other $W$ and
$Z$ analysis channels. For the $t\bar{t}H$ analysis presented in
Section~\ref{sec:tth} this approach requires an brief motivation: we
know that all backgrounds except for the $t\bar{t}b\bar{b}$ continuum
can be targeted with global kinematic cuts~\cite{tth,buckets}. For the
irreducible continuum background we can ideally reconstruct the top
momenta using a top tagger~\cite{heptop,top_tagger}. Because we are
mostly interested in different phase space regions for the $b\bar{b}$
pair the assumption of measured top momenta is appropriate, as long as
we do not consider the strictly maximum significance a realistic
estimate.

\section{Boosted $\mathbf{ZH}$ production} 
\label{sec:zh}

The first channel for which we would like to quantify the benefits of
specific, boosted phase space regions is $ZH$ production at 14~TeV
with a Higgs decay to $b$-quarks~\cite{bdrs}. For the Higgs mass of
126~GeV the corresponding branching ratio ranges around
58\%~\cite{hdecay}. Both Higgs decay jets are $b$-tagged.  Since
$p_{T,bb}$ hardly exceeds 250~GeV for the relevant events,
approximately shared between the two tagged $b$-jets, we assume a
constant single $b$-tagging efficiency of 60\%.  

For our statistical analysis we assume the $Z$ decays to $\ell =
e,\mu$ to be reconstructed perfectly. Because all the leading
backgrounds also include this $Z$ decay, possible small detector
effects will hardly impact our results.  As detector efficiencies we
include a rough factor 60\% for the lepton pair, approximately
correcting for the fact that we would probably only use leptons close
to the $Z$ pole and that not all leptons end up in the central
detector with $p_{T,\ell}>10$~GeV and $|\eta_\ell|<2.5$. These global
efficiencies mainly ensure that our integrated maximum significance is
not completely unrealistic; they hardly impact our study of the phase
space distribution of this significance.  For the signal event
generation we replace the Breit--Wigner shape of $m_{bb}$ by a
Gaussian with the experimental resolution of $\pm 12$~GeV. The
strictly speaking appropriate convolution of the Breit--Wigner shape
with the physical Higgs width and the Gaussian shape based on the
experimental resolution is very well approximated by the Gaussian
alone~\cite{orig}. Higher--order corrections to the $ZH$ production
rate~\cite{zh_nnlo} are included as a variable global scaling factor
of the signal rate.\medskip

\begin{figure}[t]
\includegraphics[height=0.31\textwidth]{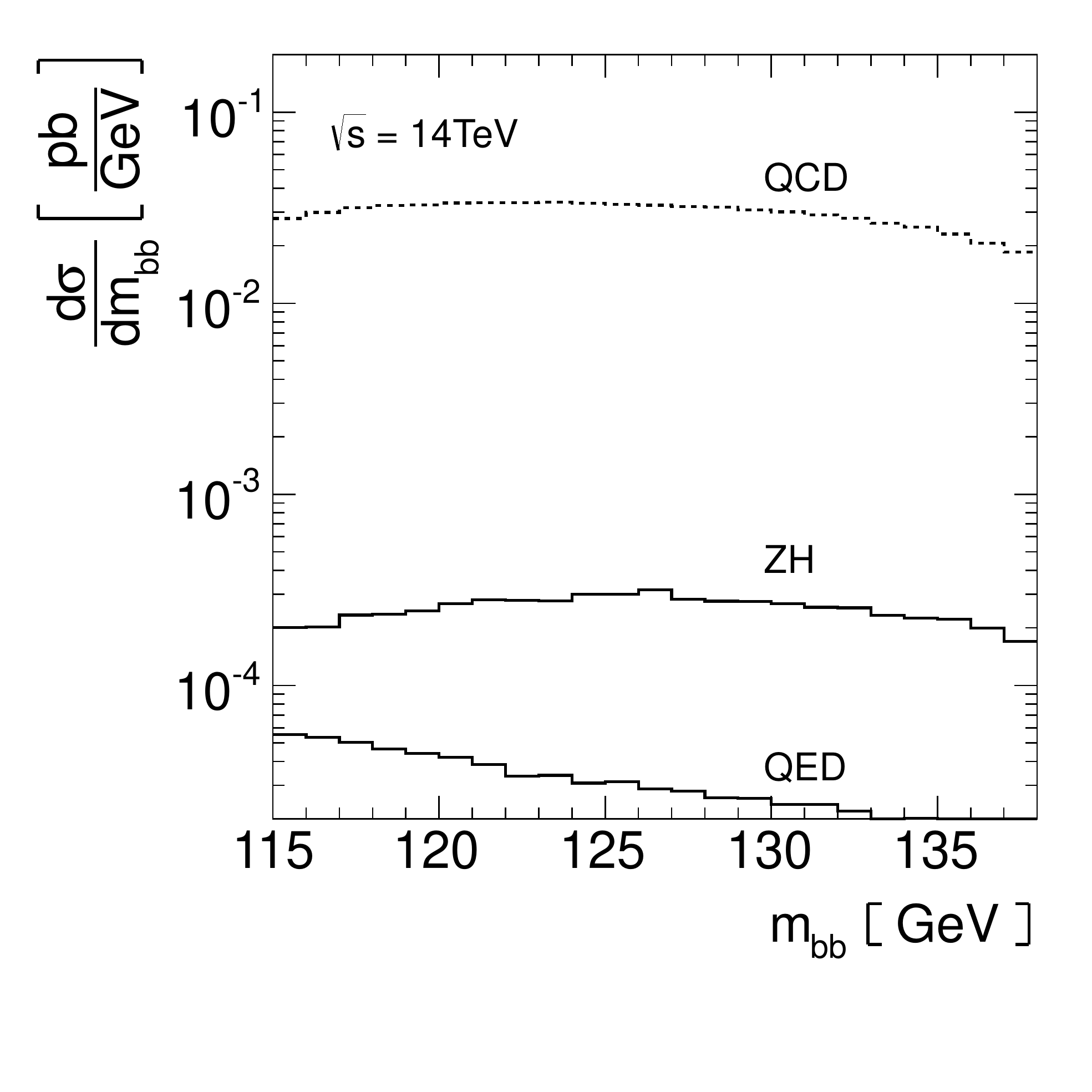}
\hspace*{0.01\textwidth}
\includegraphics[height=0.31\textwidth]{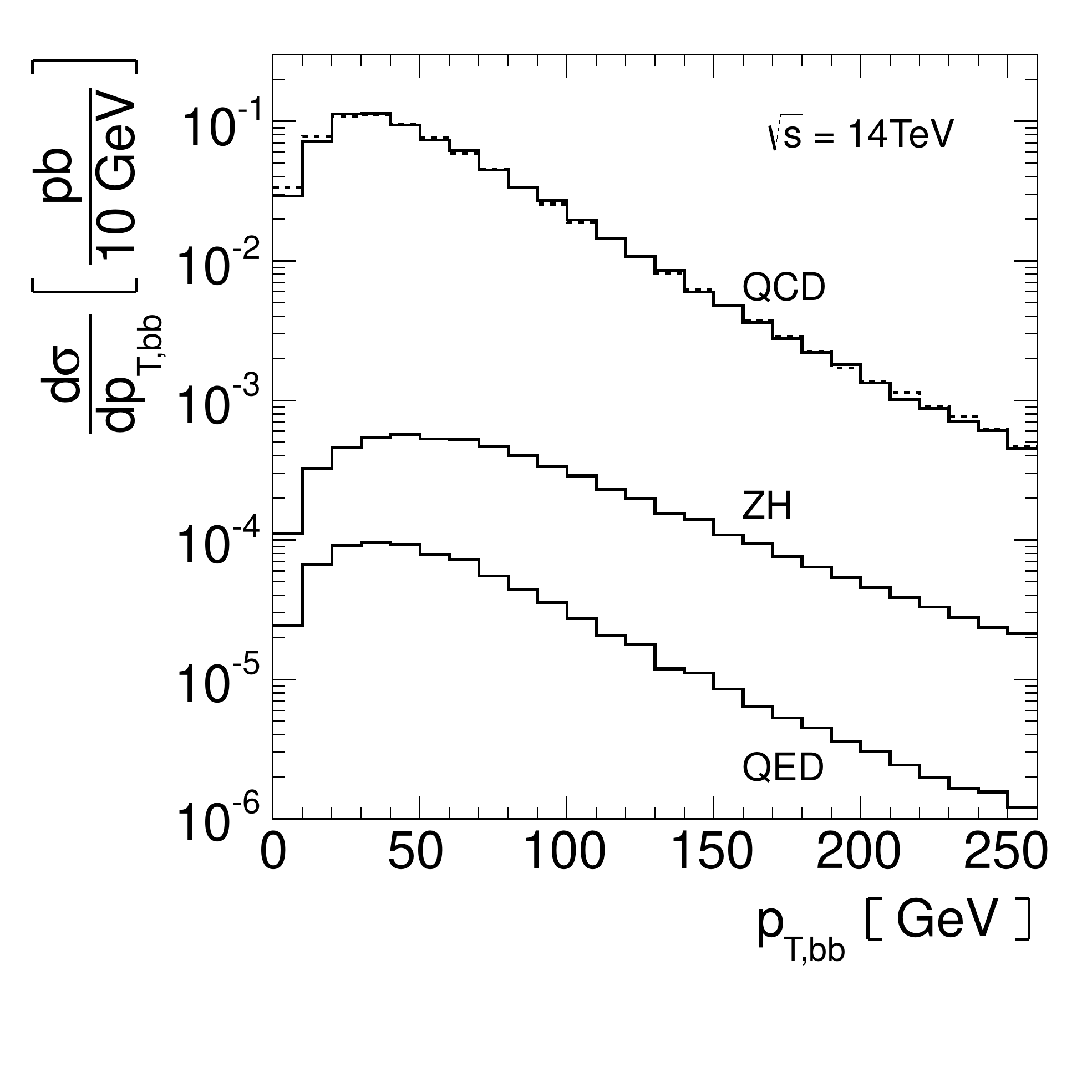}
\hspace*{0.01\textwidth}
\includegraphics[height=0.31\textwidth]{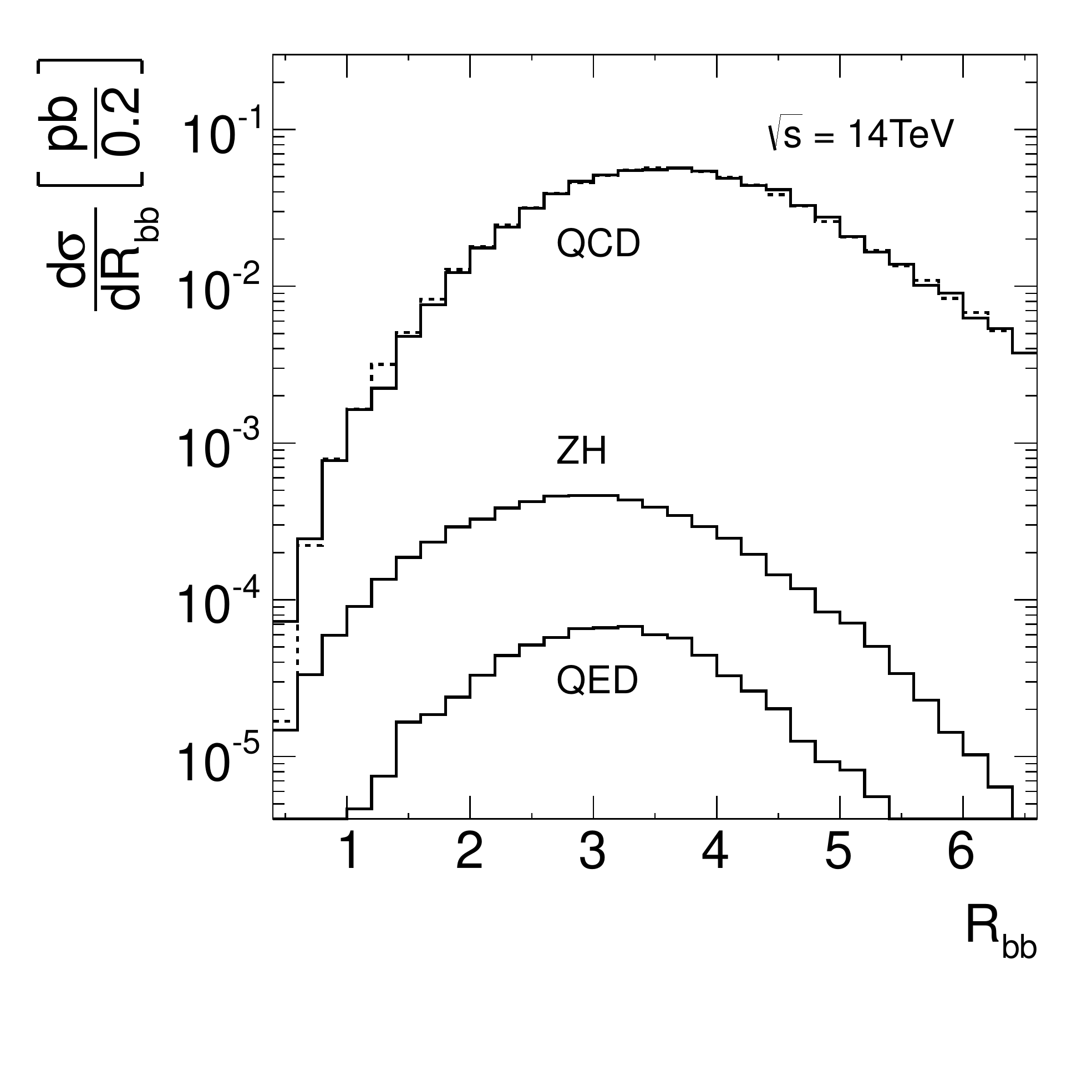} \\[-7mm]
\caption{Left: $m_{bb}$ of signal and backgrounds in the range we are
  considering. Center: transverse momentum of the $b\bar{b}$ system
  for $m_{bb} = 114 - 138$~GeV.  For the QCD background the solid line
  approximates the merged results by using the $p_{T,bb}$-dependent
  correction factor of Eq.\eqref{eq:zh_kfac}.  Right: angular
  separation $R_{bb}$ for the signal and background, including the
  $p_{T,bb}$ correction for the QCD background. The merged multi--jet
  simulation are shown as dotted lines.}
\label{fig:zh_production}
\end{figure}

The main background is continuum $Zb\bar{b}$ production at the
(leading) order $\alpha \alpha_s$. A second background is the same
final state at (leading) order $\alpha^2$, which includes $ZZ$
production with one decay $Z \to b\bar{b}$.  Fake-$b$ backgrounds are
negligible in comparison and have no unique phase space features
which would force us to consider them beyond a correction to the
$b$-tagged continuum QCD backgrounds.  A major issue of the QCD
continuum background, partly related to the invariant mass of the
$b$-jets, is the poor convergence of the total cross section as a
series in $\alpha_s$. First, to avoid issues with gluon splitting into
two $b$-quarks versus $t$-channel production of two widely separated
$b$-jets we require a mass window of $m_{bb} = 114 - 138$~GeV all
through our analysis.\footnote{It is well known that control regions
  where the effective hard process is $Zg^* \to Zb\bar{b}$ production
  should not be used to probe QCD features of continuum $Zb\bar{b}$
  production.}  In terms of the maximum likelihood this mass window
might appear overly conservative, but on the other hand we expect
experimental analyses to apply such a window to define clear side
bands. Technically, it is trivial to extend this mass window to two or
three standard deviation within \textsc{MadMax}.  For the signal our
window captures 68\% of the total cross section. We show the
corresponding $m_{bb}$ distributions in the left panel of
Fig.~\ref{fig:zh_production}, illustrating a rather depressing
signal--to--background ratio.\medskip

To simplify the \textsc{MadMax} analysis we generate events for the
QCD continuum background using the irreducible $Zb\bar{b}$ process.
To get a handle on the perturbative accuracy of this simulation we
also compute a $Zb\bar{b}$ event sample with up to two hard additional
jets, consistently combined between the hard matrix element and the
parton shower using \textsc{Madgraph}~\cite{madgraph} with
\textsc{Mlm} multi--jet merging~\cite{mlm}. While such an event sample
is not formally improved in fixed--order perturbation theory, it
should capture the leading effects from large logarithms as well as
from initial states opening only in combination with additional jets
in the final state~\cite{zbb_nlo}.  The difference between the total
leading--order $Zb\bar{b}$ rate to the merged prediction implies
correction factors around 2.1.  For the continuum QCD background the
distribution of the merged sample including up to two hard jets is
indeed harder than for the fixed--order $Zb\bar{b}$ process. While we
use the simpler, fixed--order event sample in our \textsc{MadMax}
analysis, we reweight it to the merged $p_{T,bb}$ distribution using
the $p_{T,bb}$-dependent correction factor
\begin{alignat}{5}
\log \frac{d\sigma_\text{ME+PS}}{d\sigma_\text{LO}} = 
0.65 + 1.1\times 10^{-3}\,p_{T,bb} 
     + 4.0\times 10^{-6}\,p_{T,bb}^2  \; .
\label{eq:zh_kfac}
\end{alignat}
The increasing form as a function of $p_{T,bb}$ we limit by fixing the
cross section ratio for all values above $p_{T,bb}>350$~GeV to the
maximum value, even though the number of events in this phase space is
too small to observe any effects from such a cut-off.

We illustrate the $p_{T,bb}$ distributions for the signal and the
backgrounds in the center panel Fig.~\ref{fig:zh_production}.  The
$p_{T,H}$ distribution for the signal is slightly less steep than the
corresponding background distributions, owing to the gluon parton
densities and a general QCD preference for small invariant masses
between jets. Nevertheless, the Poisson factor in our statistical
analysis will essentially remove the few events with $p_{T,H} >
250$~GeV for an integrated luminosity in the $50 - 100~\ifb$ range.
As a test, we also show the geometric separation of the two $b$-jets
in the right panel of Fig.~\ref{fig:zh_production}. The
$p_{T,bb}$-dependent correction perfectly reproduces the multi-jet
merged distribution for this observable.\medskip

\begin{figure}[t]
\includegraphics[width=0.31\textwidth]{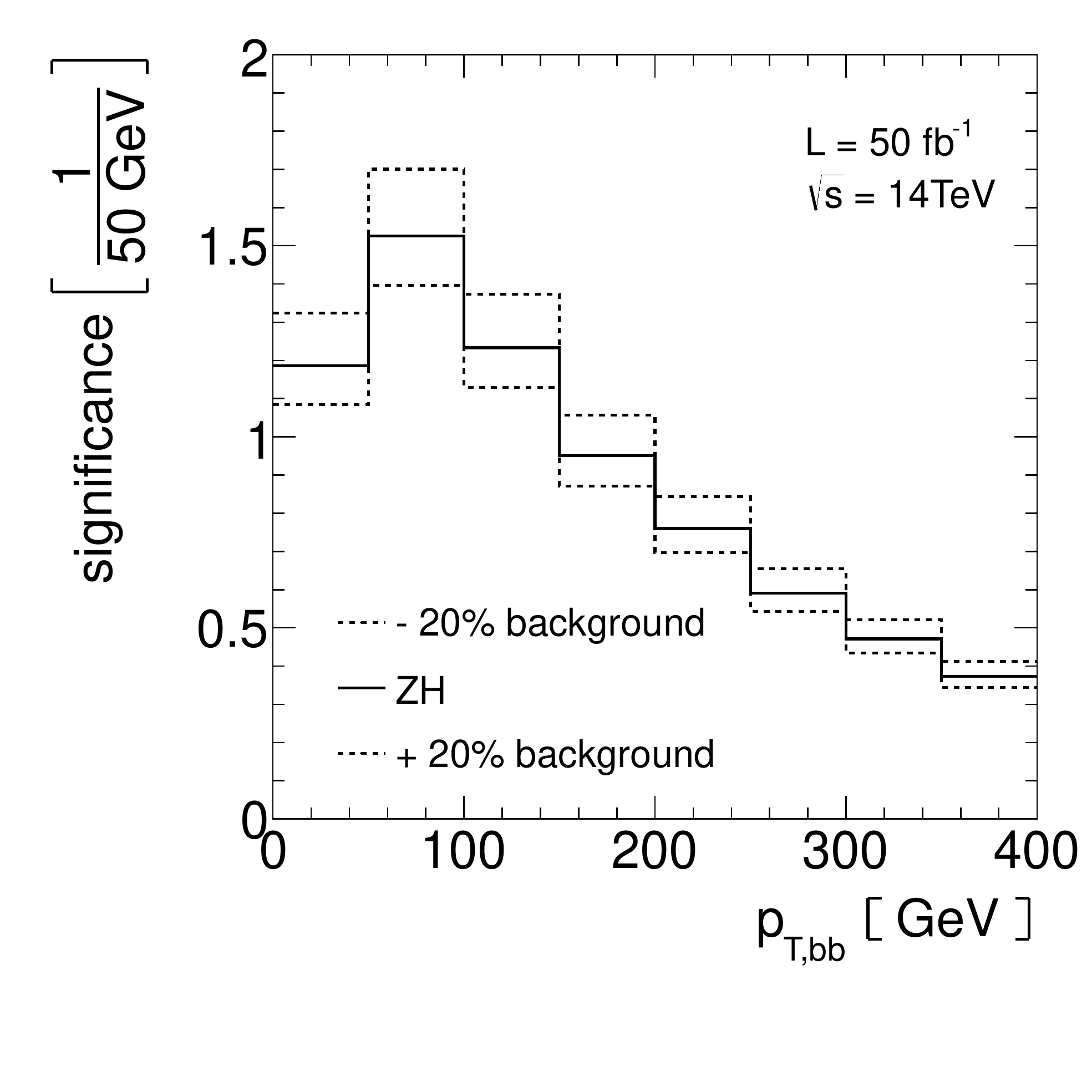}
\hspace*{0.1\textwidth}
\includegraphics[width=0.31\textwidth]{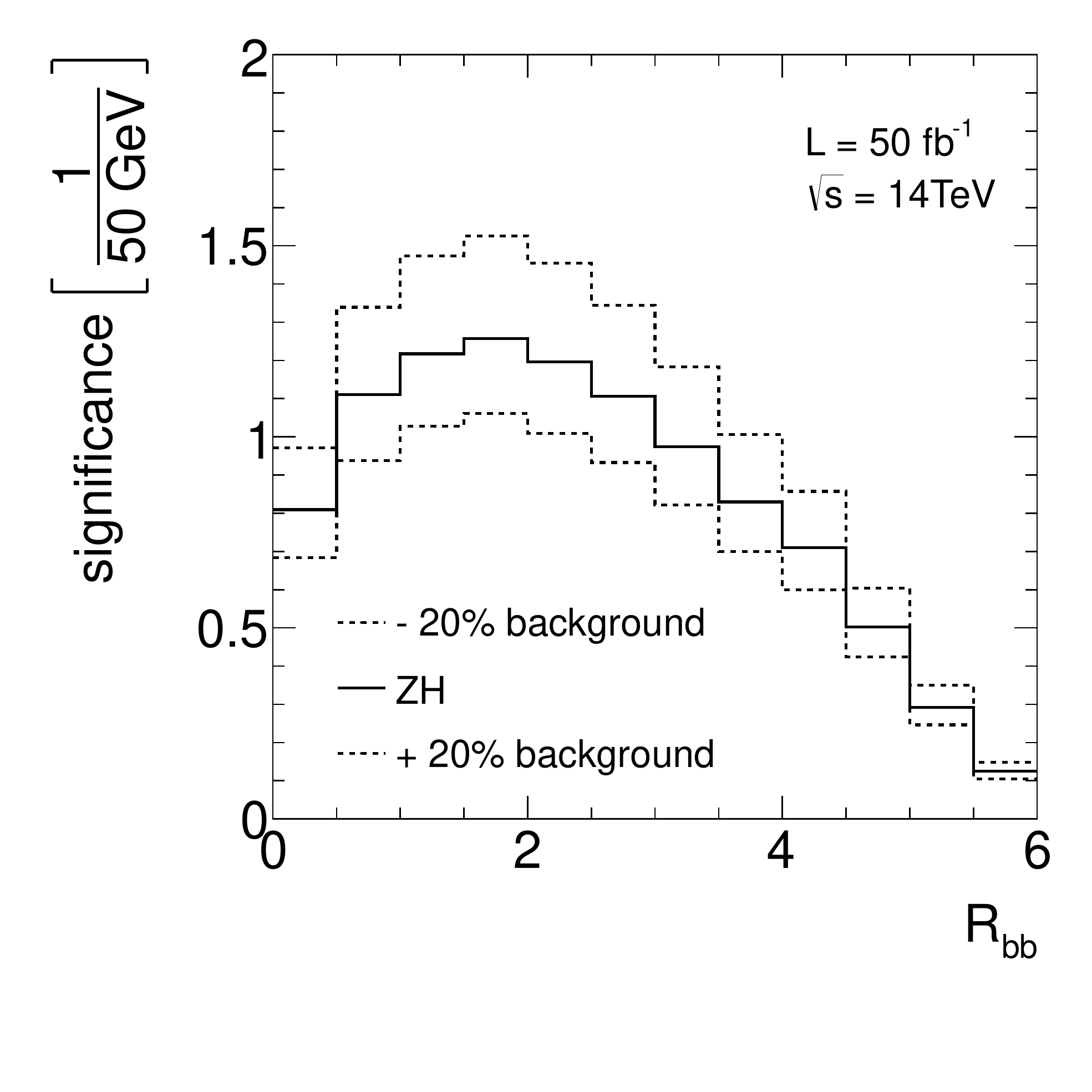} \\[-8mm]
\caption{Maximum significance for the $ZH$ signal for slices in the
  reconstructed $p_{T,bb}$ (left) and the geometric separation of the
  $b$-jets, $R_{bb}$ (right).  We only consider events inside
  the mass window $m_{bb} = 114 - 138$~GeV. The significance is
  computed for an integrated luminosity of $50~\ifb$.}
\label{fig:zh_slices}
\end{figure}

As an estimate of the maximum significance over the entire phase space
we obtain $2.7 \pm 0.3~\sigma$ for an integrated luminosity of
$50~\ifb$. The error bar is given by a $\pm 20\%$ variation in the
normalization of the signal rate. The total maximum significance at
higher integrated luminosities can be approximately computed by a
Gaussian scaling. In the left panel of Fig.~\ref{fig:zh_slices} we
show how this maximum significance is shared between slices of
$p_{T,bb}$. The phase space regime best suited to distinguish signal
from background events at this limited luminosity is
\begin{alignat}{5}
p_{T,H} = 50 - 100~\gev \; .
\end{alignat}
Boosted Higgs analyses in the $ZH$ channel indeed significantly reduce
the QCD continuum background, but Higgs taggers should be
optimized for as low $p_{T,H}$ values as possible.  The reason for
this finite transverse momentum range is on the one hand that for the
background there does not exist a large mass scale in the process,
which means that a sizeable $b\bar{b}$ invariant mass has to be
generated through a geometric separation, namely $m_{bb}^2 \simeq 2E_1
E_2 (1 - \cos \theta)$. In contrast, if we require the $H \to
b\bar{b}$ decay to be boosted, the $b$-jets in the signal will move
closer together, making it harder for the QCD background to fake the
Higgs signal. On the other hand, while we would naively expect higher
transverse momenta to carry more and more weight in the analysis, the
number of signal events in this range is strongly limited. Asking for
$p_{T,bb} > 150$ already drives us into a strongly statistics limited
phase space regime.

In the right panel of Fig.~\ref{fig:zh_slices} we show how the maximum
significance is composed by slices in the geometric separation
$R_{bb}$. This example illustrates how \textsc{MadMax} can be used to
analyze the maximum significance distribution in terms of any phase
space observable. The separation of the two $b$-tagged jets is crucial
for the definition of the fat jet as the starting point of any Higgs
tagger.  From Fig.~\ref{fig:zh_production} it is clear that hardly any signal
events lie in the range $R_{bb} \lesssim 1.0$. To optimize a boosted
Higgs analysis it appears to be beneficial to extend the fat jet size
towards $R_{bb} \sim 2.0$.\medskip

Additional effects can still modify the outcome of our study. First, we
assume that the detector performance does not depend on the boost of
the $b\bar{b}$ system. This is clearly only true up to a certain
$p_{T,bb}$ range, where the two bottom jets start overlapping. Second,
we assume theory uncertainties to be independent of $p_{T,bb}$. It is
not clear if this statement holds for QCD effects, and it is clearly
not true in the electroweak sector, once we include electroweak
Sudakov logarithms. Finally, we did not actually study dangerous
observables, for example, defined in the pre-jet stage of the analysis
or at odds with the assumed factorization properties of the signal and
background predictions. In that sense \textsc{MadMax} clearly does not
deliver a realistic estimate of systematic and theoretical
uncertainties. It is merely a first step which allows us to study
phase space patterns easily and reliably. For example the question to
what degree the estimated uncertainties are realistic and what the
effect of shape uncertainties in the background might be will be left
to possible further studies.

\section{Boosted $\mathbf{t\bar{t}H}$ production} 
\label{sec:tth}

The second process for which we want to ask the question where the
main distinguishing phase space features exist is $t\bar{t}H$
production with a Higgs decay to $b$-quarks. Two studies indicate that
boosted top and Higgs configurations might be promising to extract the
signal from the background: for purely hadronic events the buckets
methods successfully targets moderate transverse momenta around
$p_{T,t} \gtrsim 150$~GeV~\cite{buckets}. For semileptonic top pairs
the \textsc{HEPTopTagger} study shows that slightly larger boosts
$p_{T,t} \gtrsim 200 - 250$~GeV can be successfully
probed~\cite{tth}. Purely leptonic top pairs have recently been shown
to lead to promising results based on a \textsc{MadWeight}
study~\cite{tth_lep}.  In the semi--leptonic and hadronic cases it is
obvious that the boosted kinematics is an excellent way to resolve
combinatorial issues~\cite{tth,tth_comb}. The open question is to what
degree the arguments presented in the last section also point to a
boosted $t\bar{t}H$ search in terms of the signal and background
matrix elements.\medskip

To answer this question we will first assume that the continuum
$t\bar{t}b\bar{b}$ background is the most relevant issue. For the
semileptonic analysis this has been shown, once we require at least
three $b$-tags~\cite{tth}. For the purely hadronic channel the removal
of the QCD backgrounds is considerably more tedious, but appears to be
possible~\cite{buckets}. We follow this study and as a first step
require four $b$-tagged jets with an efficiency of 60\% each. In
addition, we assume a set of global cuts or other ways to reduce the
multi--jet backgrounds with a conservatively estimated efficiency
around 10\% for the $t\bar{t}H$ signal and the irreducible
$t\bar{t}b\bar{b}$ background.

To study the combinatorics of the $b$-jets identified as Higgs decay
we would have to simulate top decays. However, this is precisely the
issue which we want to separate from our significance study over phase
space. Therefore, we can assume the top quarks to be fully
reconstructed.  For hadronic top decays a major part of the
buckets~\cite{buckets} and \textsc{HEPTopTagger} studies~\cite{heptop}
have been devoted to quantifying the quality of this momentum
reconstruction.  Moreover, in the absence of additional missing energy
from the hard process one hadronic top tag can be combined with an
efficient approximate reconstruction of the boosted leptonic top
decay~\cite{leptonic_top}.  The efficiency of actually reconstructing
a hadronic top decay using a top tagger is strongly dependent on the
transverse momentum. We roughly estimate it to 33\% per top quark, in
addition to the branching ratio of 68\%. This way we should obtain at
least a semi-realistic number for the integrated maximum
significance. As mentioned before, such global efficiencies will not
affect the main outcome of our study, \ie the distribution of the
maximum significance over the transverse momentum range of the heavy
particles.  

For the Higgs decay to bottoms we again include a branching ratio of
58\%~\cite{hdecay}.  To approximately include detector effects we
evaluate the Higgs propagator with a Gaussian width of $\pm
12$~GeV. The mass window of $m_{bb} = 114-138$~GeV with a signal
efficiency of 68\% we carry through the entire analysis, both for the
signal and for the background. As discussed in the previous section
this ensures that we avoid gluon splitting issues in the background
simulation and at the same time allows for side bands in the obvious
$m_{bb}$ distribution~\cite{tth}.\medskip

\begin{figure}[t]
\includegraphics[height=0.31\textwidth]{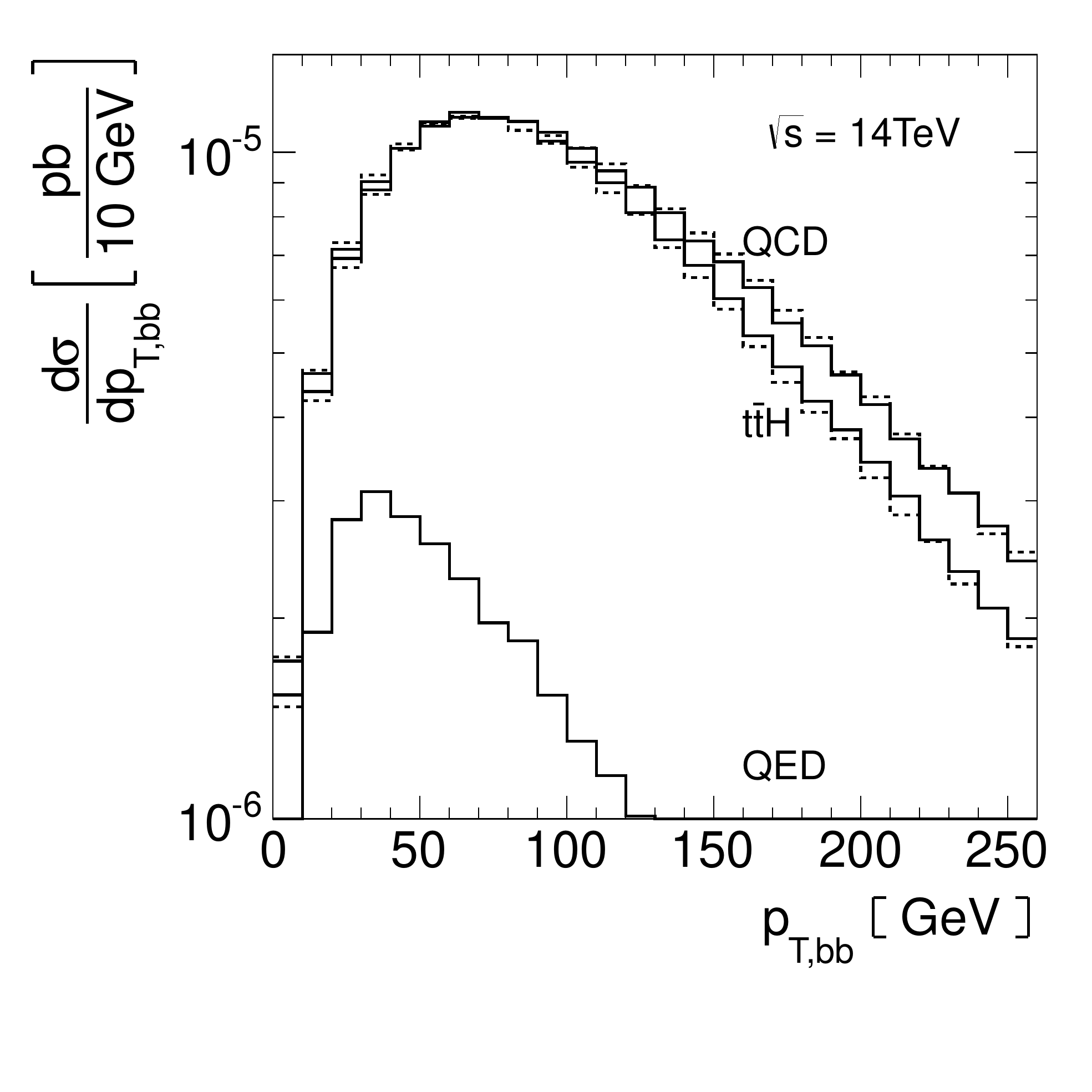}
\hspace*{0.01\textwidth}
\includegraphics[height=0.31\textwidth]{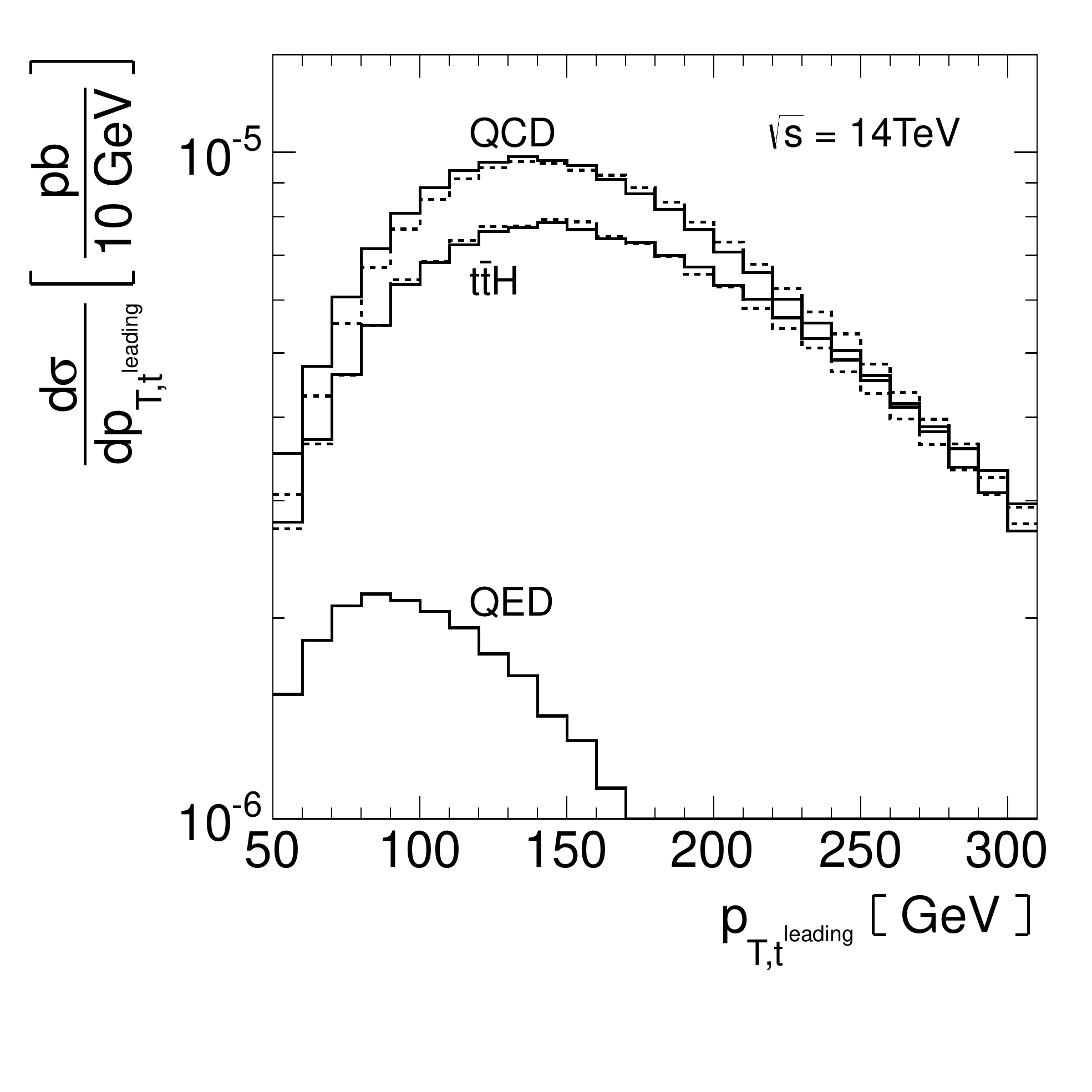}
\hspace*{0.01\textwidth}
\includegraphics[height=0.31\textwidth]{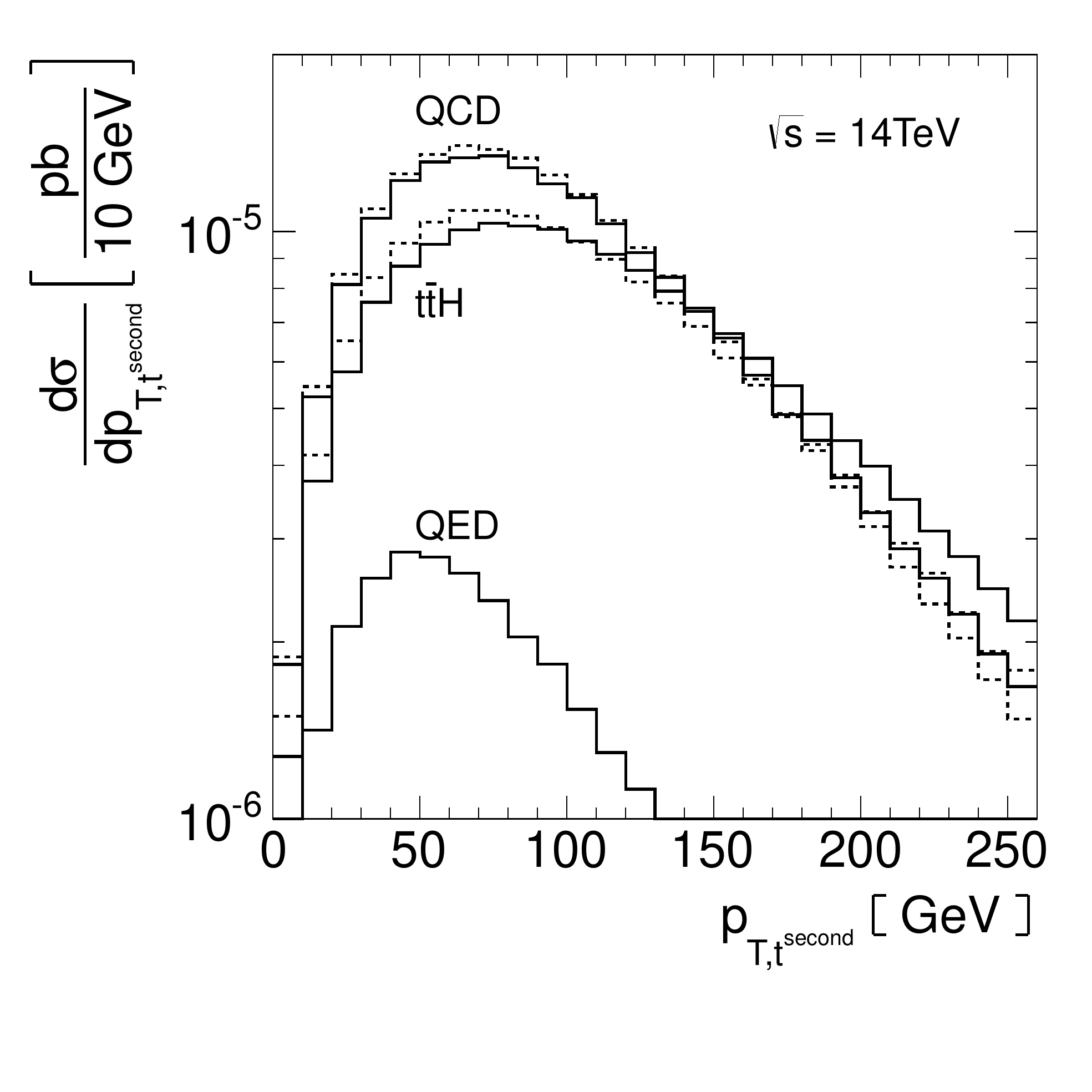} \\[-8mm]
\caption{Left: $p_{T,bb}$ for signal and backgrounds inside the mass
  window $m_{bb} = 114 - 138$~GeV. Center and right: transverse
  momenta of the two tops.  For the signal and the QCD background the
  solid lines approximate the merged results by using the
  $p_T$-dependent correction factors of Eqs.\eqref{eq:tth_kfac_s}
  and~\eqref{eq:tth_kfac_b}. The merged multi--jet simulation are
  shown as dotted lines.}
\label{fig:tth_pt}
\end{figure}

For the $t\bar{t}H$ signal the next-to-leading order corrections are
known~\cite{tth_nlo}. Since our study is focused on the phase space
structure, we approximately include them by correcting the $p_{T,H}$
distribution to agree with a matched $t\bar{t}H$ plus zero and one jet
simulation in the \textsc{Mlm} scheme~\cite{mlm} the same way as we do
with the $Zb\bar{b}$ background in Eq.\eqref{eq:zh_kfac}. We find a
correction factor for the $t\bar{t}H$ signal,
\begin{alignat}{5}
\log \frac{d\sigma_\text{ME+PS}}{d\sigma_\text{LO}} = 
0.53 - 2.5\times 10^{-3}\,p_{T,H} 
     + 2.0\times 10^{-5}\,p_{T,H}^2
     - 3.9\times 10^{-8}\,p_{T,H}^3 \; .
\label{eq:tth_kfac_s}
\end{alignat}
For a $t\bar{t}b\bar{b}$ background study in the boosted Higgs regime
it is absolutely crucial that we correctly simulate the transverse
momentum of the $b\bar{b}$ system. While the events entering the
\textsc{MadMax} analysis are generated as $t\bar{t}b\bar{b}$
production in \textsc{Madgraph}, we can correct the $p_{T,bb}$
distribution to agree with matched $t\bar{t}b\bar{b}$ plus zero and
one jet simulation using
\begin{alignat}{5}
\log \frac{d\sigma_\text{ME+PS}}{d\sigma_\text{LO}} = 
0.98 - 6.7\times 10^{-3}\,p_{T,bb} 
     + 3.8\times 10^{-5}\,p_{T,bb}^2
     - 7.6\times 10^{-8}\,p_{T,bb}^3 \; .
\label{eq:tth_kfac_b}
\end{alignat}
As for the $ZH$ case we cut off both correction factors using a
constant value above $p_{T,bb} = 350$~GeV. This reweighting of the
differential cross section should account for the leading logarithmic
higher--order corrections~\cite{tth_nlo,ttbb_nlo}, in particular
linked to different partons in the initial state.  In the left panel
of Fig.~\ref{fig:tth_pt} we show the corresponding distributions,
after including the two $p_T$-dependent correction factors.\medskip

\begin{figure}[t]
\includegraphics[width=0.31\textwidth]{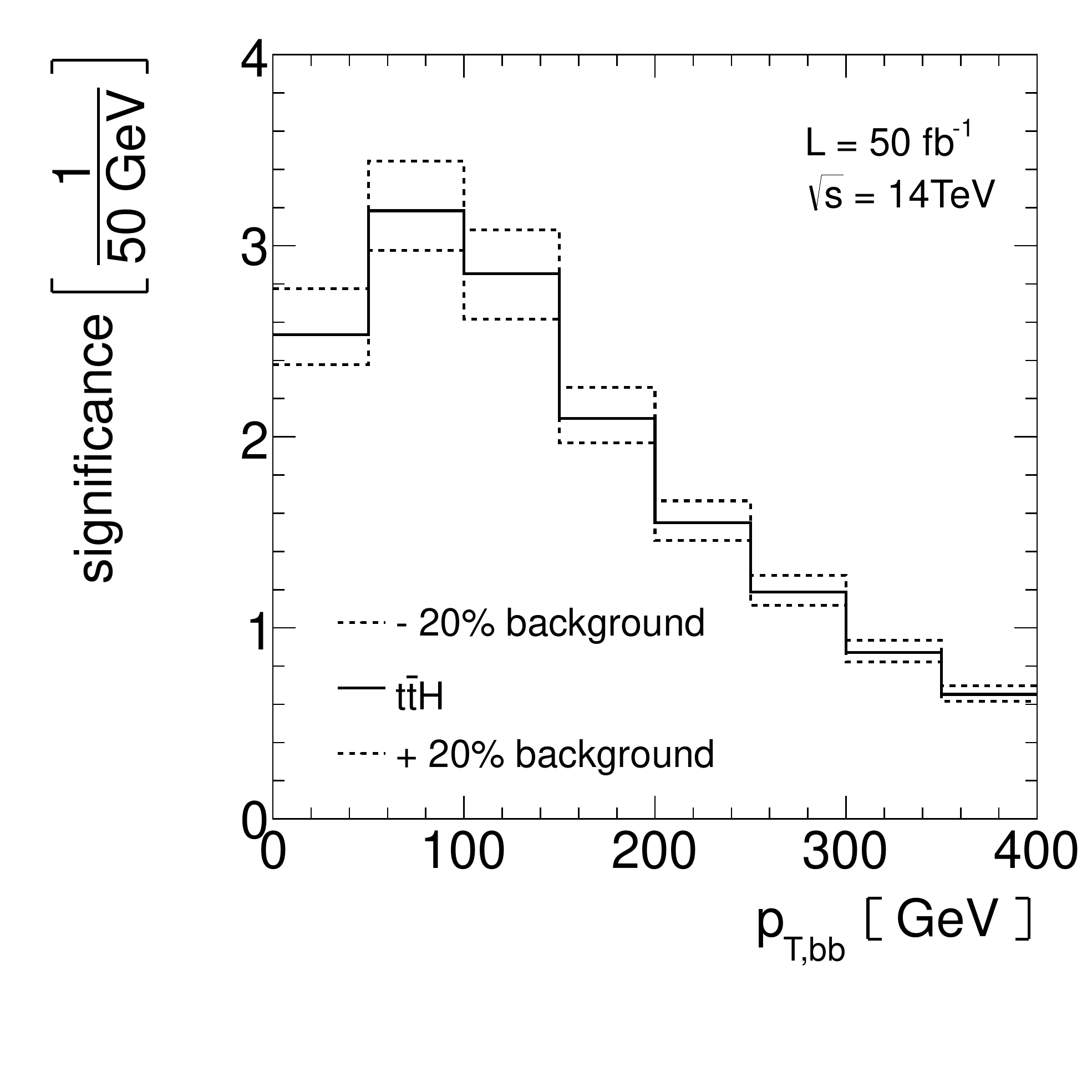}
\hspace*{0.01\textwidth}
\includegraphics[width=0.31\textwidth]{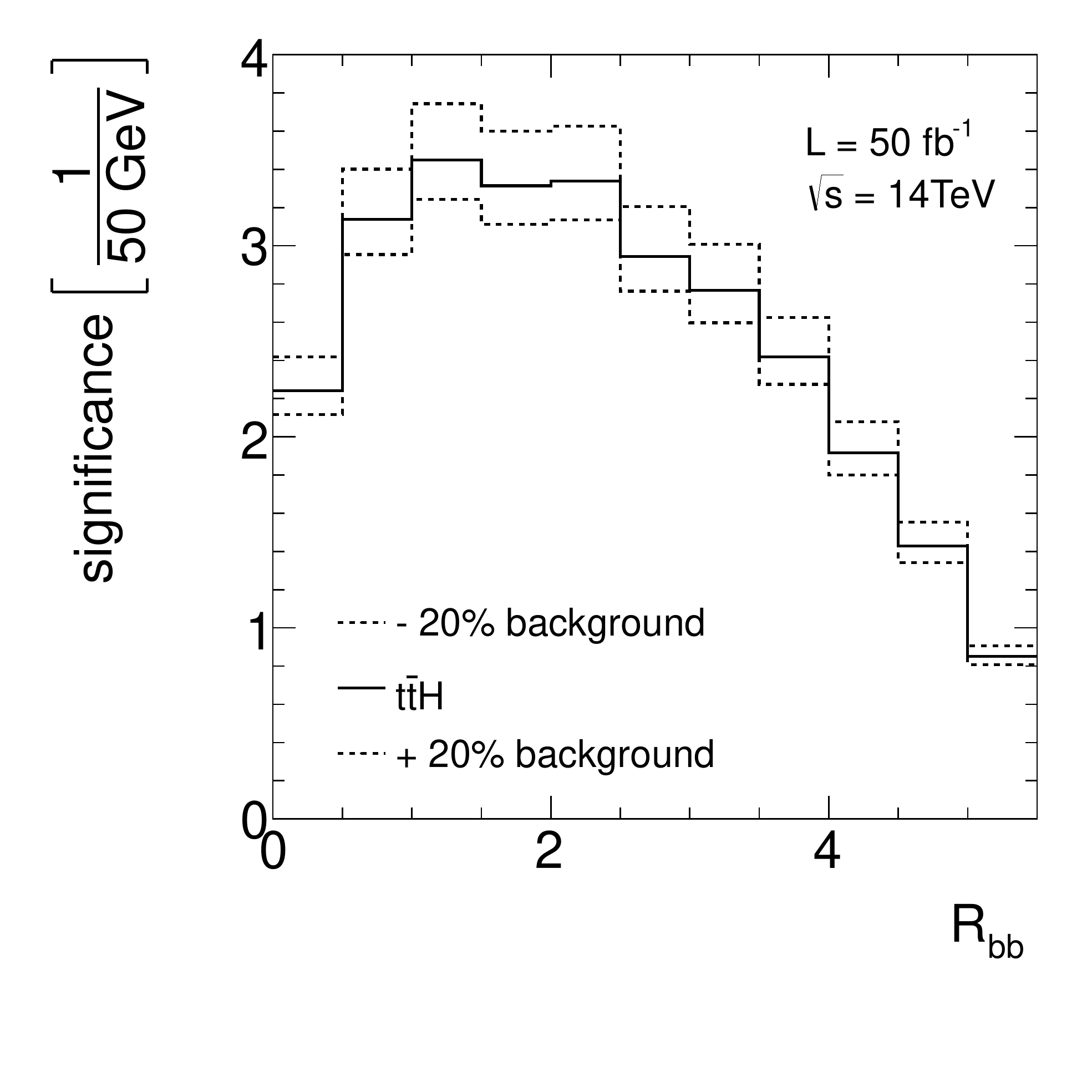}
\hspace*{0.01\textwidth}
\includegraphics[width=0.31\textwidth]{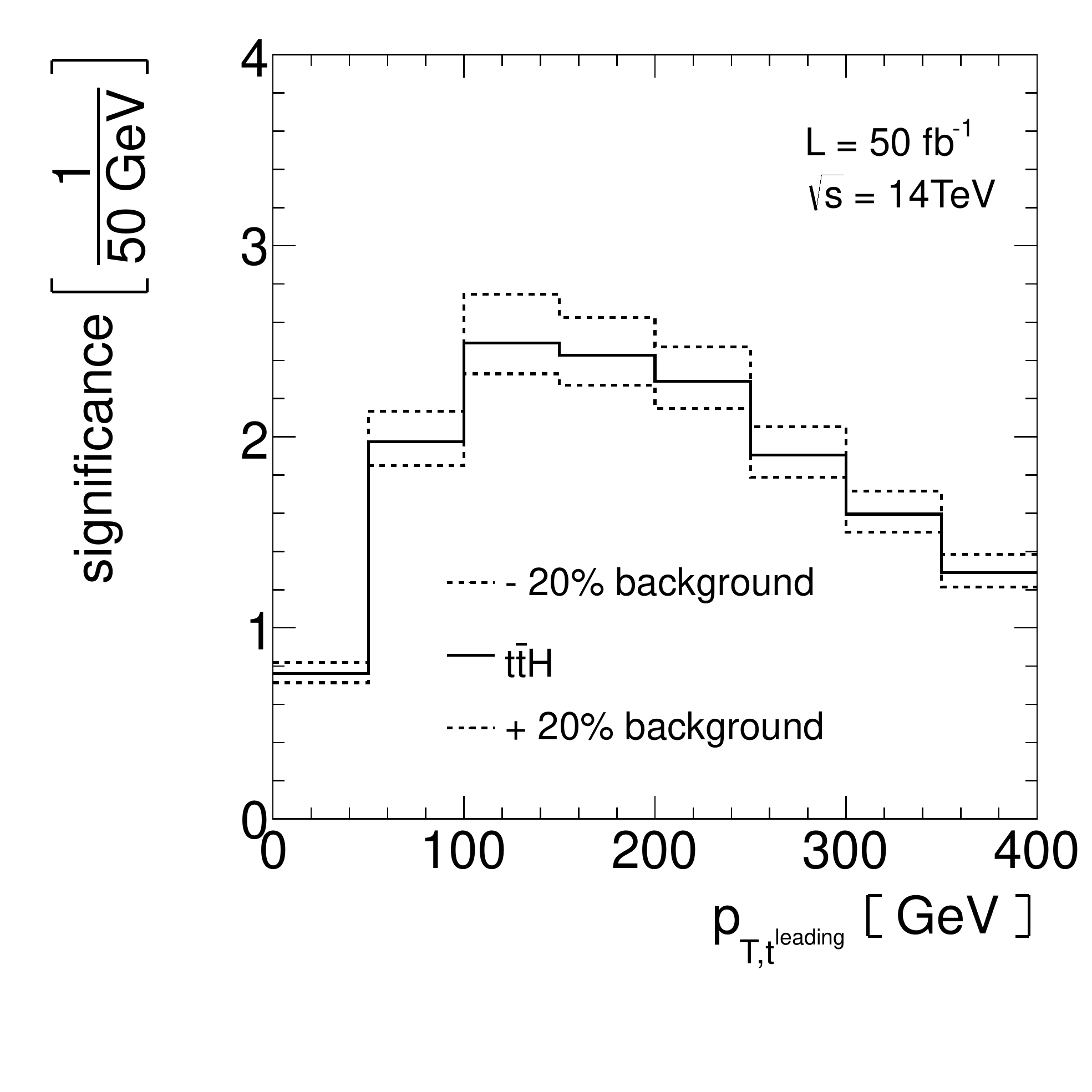} \\[-8mm]
\caption{Maximum significance for the $t\bar{t}H$ signal for slices in
  the reconstructed $p_{T,bb}$ (left), the geometric separation of the
  two $b$-jets, $R_{bb}$ (center), and the leading top
  transverse momentum $p_{T,t^\text{leading}}$ (right).  We only
  consider events inside the mass window $m_{bb} = 114 - 138$~GeV. The
  significance is computed for an integrated luminosity of $50~\ifb$.}
\label{fig:tth_slices}
\end{figure}

For the $t\bar{t}H$ analysis not only the boost of the Higgs
candidate, but also the boost of each top is relevant for the
analyses~\cite{tth,buckets}. Therefore, we need to test to what degree
the reweighting in Eqs.\eqref{eq:tth_kfac_s} and~\eqref{eq:tth_kfac_b}
affects the top kinematics. In the center and right panels of
Fig.~\ref{fig:tth_pt} we compare the background $p_{T,t}$
distributions for the fully merged event sample with the
$p_{T,bb}$-reweighted $t\bar{t}b\bar{b}$ sample. We see that the only
phase space region not perfectly described by the $p_{T,bb}$-dependent
correction factor is the low-$p_T$ range of the two tops. However, the
difference is a mere 5\%, covered by the theory uncertainty, in a
phase space region which will turn out relatively unimportant.\medskip

For the set of approximate efficiencies listed above we estimate the
maximum significance of the $t\bar{t}H$ to $5.3 \pm 0.5~\sigma$ for an
integrated luminosity of $50~\ifb$ at a collider energy of 14~TeV.
The quoted error bar is again defined by a $\pm 20\%$ variation of the
signal rate. Again, values for higher luminosities can be inferred by
Gaussian scaling. In Fig.\ref{fig:tth_slices} we show this maximum
significance in slices of the transverse momentum of the reconstructed
Higgs system, the separation of the Higgs decay jets, and the
transverse momentum of the leading top. For the Higgs boost we find a
similar range as for the $ZH$ channel, while the most promising
transverse momentum range for the heavier top roughly scales with
$m_t/m_H$,
\begin{alignat}{5}
p_{T,H} &= \; 50 - 100~\gev \notag \\
p_{T,t} &= 100 - 250~\gev \; .
\end{alignat}
This result indicates that the requirements for the Higgs
tagger~\cite{bdrs} in the $ZH$ and $t\bar{t}H$ analyses are very
similar. The only difference between these two channels is that for
the $t\bar{t}H$ process the Higgs tagger has to be adapted to higher
jet multiplicities, as done in Ref.~\cite{tth}.  For the top tagger it
is crucial that we gain access to transverse momenta well below
$p_{T,t} = 300$~GeV, ruling out most of the currently available
analysis tools optimized for heavy $t\bar{t}$ resonance
searches~\cite{top_tagger}.

\section{Conclusions} 
\label{sec:conclusions}

In this paper we have introduced \textsc{MadMax} as a novel approach
to studying the composition of a signal significance in terms of the
signal and background phase space. It allows us to determine those
phase regions which are best suited for the extraction of a signal
process from irreducible backgrounds in an efficient and
mathematically well--defined manner~\cite{orig}.\medskip

Relying on \textsc{MadMax} we have studied the two Higgs search
channels involving a hadronic $H \to b\bar{b}$ decay, namely
associated $ZH$ and $t\bar{t}H$ production. In both cases the central
question is to what degree boosted Higgs configurations benefit the
signal extraction and what range in transverse Higgs momenta we should
target. Unlike in the original study~\cite{orig}, we specifically did
not focus on predicting the integrated maximum significance for each
of these processes, so detector effects as well as global efficiencies
are only adjusted to obtain a semi-realistic number.

For the $ZH$ channel we find that for integrated luminosities in the
$50-100~\ifb$ range the most promising phase space regime is around or
below $p_{T,H} =100$~GeV, challenging the development of Higgs
taggers. For the $t\bar{t}H$ analysis the first reason to rely on
boosted top and Higgs decay topologies is the otherwise overwhelming
combinatorial background. In this study we estimated the additional
motivation for using these topologies based on the matrix element
structure of the signal and the irreducible background. The most
promising range in the transverse momentum of the reconstructed Higgs
boson again came out as $p_{T,H} = 50 - 100$~GeV, indicating that the
same Higgs tagger should suit both analyses. To include as many of the
relevant signal events as possible the size of the fat jet could be extended
towards $R_{bb} \sim 2$, if possible.  For the transverse momentum of
the leading top quark the significance is mostly collected for
$p_{T,t} = 100 - 250$~GeV, seriously challenging the development of
hadronic top taggers.\medskip

We note that \textsc{MadMax} is an automized tool which can be used in
the \textsc{Madgraph5} framework to produce multi--dimensional
differential distributions adding to the maximum significance in any
kinematic observable.  Some of its current limitations, like the focus
on irreducible backgrounds can be overcome
easily. Likewise, simple transfer functions can be included in a
straightforward manner. Once these effects are taken into account, we
should be able to also give a more reliable estimate of the integrated
maximum significance for a signal--background combination.\medskip

\begin{center}
{\bf Acknowledgments}
\end{center}

First of all, we would like to thank Kyle Cranmer for supporting this
extension of the original work of Ref.\cite{orig}.  PS would like to
thank the IMPRS {\sl Precision studies of fundamental symmetries} for
their never--ending support. TP would like to thank the CCPP at New
York University for their hospitality in a crucial phase of this
paper. DW is grateful to the University of Pittsburgh and to Ayres
Freitas for supporting him while writing this paper, as well as to Jan
Pawlowski for agreeing to co-referee his rather technical
thesis. Finally, all of us would like to thank Fabio Maltoni for
suggesting the name \textsc{MadMax}.

\appendix
\section{MadMax in Madgraph}
\label{sec:appendix}

To understand how \textsc{MadMax} works together with
\textsc{Madgraph5}~\cite{madgraph} we first describe some of the main
\textsc{Madgraph} features.  To construct the log--likelihood map of
Eq.\eqref{eq:likelimap} we need the squared matrix elements for the
signal and background, which \textsc{Madgraph5} computes using a
modified version of \textsc{Helas}~\cite{helas}.  These matrix
elements are integrated using the so-called single diagram enhanced
method~\cite{madevent} to account for the propagator structure. A good
example process is $pp \to \mu^+ \mu^-$, where we compute the
cross--section via
\begin{alignat}{5}
\sigma_\text{tot} = \int dx_1 \, dx_2 \, d\text{LIPS} \;
f_p(x_1,\mu_F) \, f_p(x_2,\mu_F) \; 
\left| \sum \limits_n \mathcal{M}_n \right|^2 \; .
\end{alignat}
The phase space is described by a random number vector $r$ and
transforms the integral into a sum over phase space cells $\Delta
r$. The parton densities are evaluated together with the matrix
elements, so in the following we implicitly include them in
$\mathcal{M}$.  We can improve the convergence if we know the leading
behavior of the individual matrix elements,
\begin{alignat}{5}
\sigma_\text{tot} 
= \sum \limits_r \; \Delta r \;
\left| \sum \limits_n \mathcal{M}_n (r) \right|^2 
= \sum \limits_{r,i} \Delta r \,
  \frac{ \left| \mathcal{M}_i (r) \right|^2 }{ \sum \limits_n
    \left| \mathcal{M}_n (r) \right|^2 } \times \left| \sum
    \limits_n \mathcal{M}_n (r) \right|^2 \; .
\end{alignat}
%
The optimized phase space mapping is different for each diagram
$i$. For our example we optimize for a $1/s$ scaling in the photon
exchange and for a Breit--Wigner propagator in the $Z$ exchange.  In
addition, the sum over diagrams is decomposed into incoherent partial
sums for different incoming partons, so an additional weight accounts
for the parton densities,
\begin{alignat}{5}
\sigma_\text{tot} = 
\sum \limits_{r,i,p} \Delta r \,
\frac{\left| \mathcal{M}_{i_p} (r) \right|^2}
     {\sum \limits_{n_p} \left| \mathcal{M}_{n_p} (r) \right|^2}
\times \left| \sum \limits_{n_p} \mathcal{M}_{n_p} (r) \right|^2 
\times \omega_p^\text{pdf} (r) \; .
\end{alignat}
\medskip

In this framework \textsc{MadMax} computes the maximum significance
using a log--likelihood ratio integration. It starts from the single
event likelihood for signal and background. For each phase space
point we need to know simultaneously 
\begin{alignat}{5}
d\sigma_s(r) &= 
\left| \sum \limits_{n_s} \mathcal{M}_{n_s}(r) \right|^2
\qqqquad
d\sigma_b (r) = 
\left| \sum \limits_{n_b} \mathcal{M}_{n_b}(r) \right|^2
\notag \\
q(r) &= -\sigma_{\text{tot,s}} \, \mathcal{L} 
   + \log \left( 1 + \frac{d\sigma_s(r)}{d\sigma_b(r)} 
          \right) \; ,
\end{alignat}
as quoted in Eq.\eqref{eq:likelimap}. From the logarithm in the
log--likelihood ratio it is clear that the we cannot use the single
enhanced diagram method.  Instead, we will use a modified version
closer to the original proposal~\cite{madevent}: let us assume our
signal process consists of $a_s \in [1, \ldots, n_s]$ sub-processes,
with different partons in the initial and final state.  Each
sub-process will be computed using $i_{a_s} \in [1 , \ldots, n_{a_s}
]$ matrix elements. The same is true for the background, namely $a_b
\in [1, \ldots, n_b]$ subprocesses with $i_{a_b} \in [1 , \ldots,
  n_{a_b}]$ matrix elements.  Any function $f(r)$ we can construct
using the basis elements
\begin{alignat}{5}
\frac{ \left| \mathcal{M}_i (r) \right|^2 }{
\sum \limits_{\substack{a_s,i_{a_s}\\a_b,i_{a_b}}} \left|
\mathcal{M}_n (r) \right|^2 } \times f(r) \; .
\end{alignat}
In this basis we can integrate all signal and background
rates. Furthermore, the points $q(r)$ defining the single event
probability have the correct weight as well.  The only remaining issue
is that events in \textsc{Madgraph5} are usually computed with a
dynamical factorization and renormalization scale choice. In
\textsc{MadMax} we add, divide, and combine matrix elements from
signal as well as background processes. In the current implementation
of \textsc{MadMax} we use fixed factorization and renormalization
scales, to simplify the interface to \textsc{Madgraph5}. The
implementation of a general scale choice would be
straightforward.\medskip

\begin{figure}[t]
\includegraphics[width=0.85\textwidth]{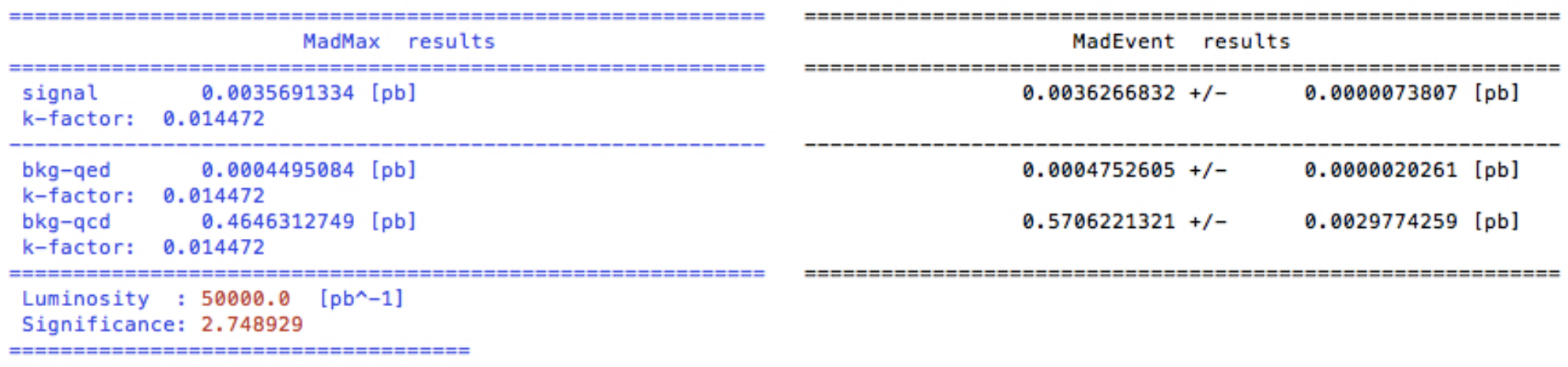}
\caption{Example \textsc{MadMax} output for the $ZH$ channel. We
  give the computed signal and background cross--sections as well as
  the corresponding standard \textsc{Madgraph} results. In addition,
  we give all constant efficiencies and $K$-factors included in the
  statistical analysis.}
\label{fig:example_results}
\end{figure}

We test and validate our implementation in two ways.  First, we have
compared our findings for weak--boson--fusion Higgs production with a
Higgs decay to muons with the original results~\cite{orig}.  Second,
for each set of signal and background processes we can compare our
signal and background cross--sections with the \textsc{Madgraph5}
results obtained in parallel, an example output is shown in
Fig.~\ref{fig:example_results}. The numbers always agree within the
numerical uncertainties. An extensive validation study as well as more
information on the implementation of MadMax can be found in
Ref.~\cite{daniel}.



\begin{thebibliography}{99}

\bibitem{orig}
 K.~Cranmer and T.~Plehn,
  Eur.\ Phys.\ J.\ C {\bf 51}, 415 (2007).

\bibitem{sfitter}
 M.~Klute, R.~Lafaye, T.~Plehn, M.~Rauch and D.~Zerwas,
  Europhys.\ Lett.\  {\bf 101}, 51001 (2013).

\bibitem{bdrs}
 J.~M.~Butterworth, A.~R.~Davison, M.~Rubin, G.~P.~Salam,
  Phys.\ Rev.\ Lett.\  {\bf 100}, 242001 (2008);
 ATLAS note,
 ATL-PHYS-PUB-2009-088.

\bibitem{tth}
 T.~Plehn, G.~P.~Salam and M.~Spannowsky,
  Phys.\ Rev.\ Lett.\  {\bf 104}, 111801 (2010).

\bibitem{buckets}
 M.~R.~Buckley, T.~Plehn and M.~Takeuchi,
  JHEP {\bf 1308}, 086 (2013);
 M.~R.~Buckley, T.~Plehn, T.~Schell and M.~Takeuchi,
  arXiv:1310.6034 [hep-ph].

\bibitem{top_tagger}
 for recent overviews see \eg
  A.~Altheimer, S.~Arora, L.~Asquith, G.~Brooijmans, J.~Butterworth, M.~Campanelli, B.~Chapleau and A.~E.~Cholakian {\it et al.},
  J.\ Phys.\ G {\bf 39}, 063001 (2012);
 T.~Plehn and M.~Spannowsky,
  J.\ Phys.\ G {\bf 39}, 083001 (2012).

\bibitem{deconstruction}
 D.~E.~Soper and M.~Spannowsky,
  Phys.\ Rev.\ D {\bf 84}, 074002 (2011);
 D.~E.~Soper and M.~Spannowsky,
  Phys.\ Rev.\ D {\bf 87}, no. 5, 054012 (2013).

\bibitem{madgraph}
 J.~Alwall, P.~Demin, S.~de Visscher, R.~Frederix, M.~Herquet, F.~Maltoni, T.~Plehn and D.~L.~Rainwater {\it et al.},
  JHEP {\bf 0709}, 028 (2007);
 J.~Alwall, M.~Herquet, F.~Maltoni, O.~Mattelaer and T.~Stelzer,
  JHEP {\bf 1106}, 128 (2011).

\bibitem{proof}
 for a proof and corresponding definitions, see \eg
 J.~Stuart, A.~Ord and S.~Arnold, 
  {\em Kendall's Advanced Theory of Statistics,
       Vol 2A (6th Ed.)} (Oxford University Press, New York, 1994); for a pedagogical introduction in the context of high-energy physics, see A. Read, in `1st Workshop on Confidence Limits', CERN Report No. CERN-2000-005 (2000).

\bibitem{early_mem}
 for some early examples see \eg
 K.~Kondo,
  J.\ Phys.\ Soc.\ Jap.\  {\bf 57}, 4126 (1988);
 D.~Atwood and A.~Soni,
  Phys.\ Rev.\ D {\bf 45}, 2405 (1992);
 M.~Diehl and O.~Nachtmann,
  Eur.\ Phys.\ J.\ C {\bf 1}, 177 (1998).

\bibitem{mem}
 see some early examples see \eg 
 V.~M.~Abazov {\it et al.}  [D0 Collaboration],
  Phys.\ Lett.\ B {\bf 617}, 1 (2005);
 A.~Abulencia {\it et al.}  [CDF Collaboration],
  Phys.\ Rev.\ D{\bf 73}, 092002 (2006);
 V.M Abazov {\it et. al.} [D0 Collaboration],  
  Nature 429, 638 (2004).
  
\bibitem{madweight}
 P.~Artoisenet, V.~Lemaitre, F.~Maltoni and O.~Mattelaer,
  JHEP {\bf 1012}, 068 (2010).

\bibitem{clfft}
 H.~Hu and J.~Nielsen, in `1st Workshop on Confidence Limits',
  CERN 2000-005 (2000) [arXiv:physics/9906010].

\bibitem{lepstats}
 K.~Cranmer, LEPStats4LHC, 
 \url{https://plone4.fnal.gov/P0/phystat/packages/0703002}

\bibitem{2hdm} 
 D.~Lopez-Val, T.~Plehn and M.~Rauch,
  JHEP {\bf 1310}, 134 (2013).

\bibitem{lecture}
 for a pedagogical introduction see \eg 
 T.~Plehn,
  Lect.\ Notes Phys.\  {\bf 844}, 1 (2012);
  [arXiv:0910.4182 [hep-ph]].

\bibitem{heptop}
 T.~Plehn, M.~Spannowsky, M.~Takeuchi and D.~Zerwas,
  JHEP {\bf 1010}, 078 (2010);
 T.~Plehn, M.~Spannowsky and M.~Takeuchi,
  Phys.\ Rev.\ D {\bf 85}, 034029 (2012);
 G.~Aad {\it et al.}  [ATLAS Collaboration],
  JHEP {\bf 1301}, 116 (2013);
  \url{http://www.physi.uni-heidelberg.de//Publications/Kasieczka-Doktor.pdf}


\bibitem{hdecay}
 A.~Djouadi, J.~Kalinowski and M.~Spira,
  Comput.\ Phys.\ Commun.\  {\bf 108}, 56 (1998).

\bibitem{zh_nnlo}
 O.~Brein, A.~Djouadi and R.~Harlander,
  Phys.\ Lett.\ B {\bf 579}, 149 (2004);
 S.~Dawson, T.~Han, W.~K.~Lai, A.~K.~Leibovich and I.~Lewis,
  Phys.\ Rev.\ D {\bf 86}, 074007 (2012).

\bibitem{mlm}
 M.~L.~Mangano, M.~Moretti, F.~Piccinini and M.~Treccani,
  JHEP {\bf 0701}, 013 (2007).

\bibitem{zbb_nlo}
 J.~M.~Campbell and R.~K.~Ellis,
  Phys.\ Rev.\ D {\bf 62}, 114012 (2000);
 F.~Febres Cordero, L.~Reina and D.~Wackeroth,
  Phys.\ Rev.\ D {\bf 80}, 034015 (2009).

\bibitem{tth_lep}
 P.~Artoisenet, P.~de Aquino, F.~Maltoni and O.~Mattelaer,
  arXiv:1304.6414 [hep-ph].

\bibitem{tth_comb}
 J.~Cammin and M.~Schumacher,
  ATL-PHYS-2003-024;
 for an experimental update see \eg 
 S.~Allwood-Spiers,
  ATL-PHYS-PROC-2011-280.

\bibitem{leptonic_top}
 J.~Thaler and L.~-T.~Wang,
  JHEP {\bf 0807}, 092 (2008);
 K.~Rehermann and B.~Tweedie,
  JHEP {\bf 1103}, 059 (2011);
 T.~Plehn, M.~Spannowsky and M.~Takeuchi,
  JHEP {\bf 1105}, 135 (2011).

\bibitem{tth_nlo}
 W.~Beenakker, S.~Dittmaier, M.~Kr\"amer, B.~Pl\"umper, M.~Spira and P.~M.~Zerwas,
  Nucl.\ Phys.\ B {\bf 653}, 151 (2003);
  S.~Dawson, C.~Jackson, L.~H.~Orr, L.~Reina and D.~Wackeroth,
  Phys.\ Rev.\ D {\bf 68}, 034022 (2003).

\bibitem{ttbb_nlo} 
 A.~Bredenstein, A.~Denner, S.~Dittmaier and S.~Pozzorini,
  Phys.\ Rev.\ Lett.\  {\bf 103}, 012002 (2009);
 A.~Bredenstein, A.~Denner, S.~Dittmaier and S.~Pozzorini,
  JHEP {\bf 1003}, 021 (2010);
 G.~Bevilacqua, M.~Czakon, C.~G.~Papadopoulos, R.~Pittau and M.~Worek,
  JHEP {\bf 0909}, 109 (2009).

\bibitem{helas}
 H.~Murayama, I.~Watanabe and K.~Hagiwara,
  KEK-91-11,
  \url{http://madgraph.kek.jp/~kanzaki/Tutorial/helas.pdf}
  
\bibitem{madevent}
 F.~Maltoni and T.~Stelzer,
  JHEP {\bf 0302}, 027 (2003).

\bibitem{daniel}
  D.~Wiegand, master thesis (2013),
  \url{http://www.thphys.uni-heidelberg.de/\textasciitilde
  plehn/includes/theses/wiegand.pdf}

\end{thebibliography}
\end{document}